\documentclass[]{aastex631}
\usepackage{amsmath}
\usepackage{graphicx}

\newcommand{\um}{$\mu$m}
\newcommand{\mic}{$\mu$m}

\begin{document}
\title{Probing the dipole of the diffuse gamma-ray background}


\author[0000-0003-2156-078X]{A. Kashlinsky}
\affiliation{Code 665, Observational Cosmology Lab, NASA Goddard Space Flight Center, 
Greenbelt, MD 20771, USA; SSAI, Lanham, MD 20770, USA; and Dept of Astronomy, University of Maryland, College Park, MD 20742}
\email{Alexander.Kashlinsky@nasa.gov} 
\author[0000-0002-2130-2513]{F. Atrio-Barandela}
\affiliation{Department of Fundamental Physics, University of Salamanca, 37008 Salamanca, Spain}
\author[0000-0002-0504-549X]{C. S. Shrader}
\affiliation{
Code 661, Astroparticle Physics Lab., NASA Goddard Space Flight Center, 
Greenbelt, MD 20771; Department of Physics, Catholic University of America, USA}

\received{Sep 7, 2023}
\revised{Sep 27, 2023}
\accepted{Sep 28, 2023}


\def\plotone#1{\centering \leavevmode
\epsfxsize=\columnwidth \epsfbox{#1}}

\def\wisk#1{\ifmmode{#1}\else{$#1$}\fi}

\def\wm2sr {Wm$^{-2}$sr$^{-1}$ }		
\def\nw2m4sr2 {nW$^2$m$^{-4}$sr$^{-2}$\ }		
\def\nwm2sr {nWm$^{-2}$sr$^{-1}$\ }		
\def\nw2m4sr {nW$^2$m$^{-4}$sr$^{-1}$\ }
\def\Ncut {$N_{\rm cut}$\ }
\def\lt     {\wisk{<}}
\def\gt     {\wisk{>}}
\def\le     {\wisk{_<\atop^=}}
\def\ge     {\wisk{_>\atop^=}}
\def\lsim   {\wisk{_<\atop^{\sim}}}
\def\gsim   {\wisk{_>\atop^{\sim}}}
\def\kms    {\wisk{{\rm ~km~s^{-1}}}}
\def\Lsun   {\wisk{{\rm L_\odot}}}
\def\Msun   {\wisk{{\rm M_\odot}}}
\def\um     { $\mu$m\ }
\def\sig    {\wisk{\sigma}}
\def\etal   {{\sl et~al.\ }}
\def\eg	    {{\it e.g.\ }}
\def\ie     {{\it i.e.\ }}
\def\bsl    {\wisk{\backslash}}
\def\by     {\wisk{\times}}
\def\cosec {\wisk{\rm cosec}}
\def\mic {\wisk{ \mu{\rm m }}}

\def\amin   {\wisk{^\prime\ }}
\def\asec   {\wisk{^{\prime\prime}\ }}
\def\cc     {\wisk{{\rm cm^{-3}\ }}}
\def\deg     {\wisk{^\circ}}
\def\ddeg   {\wisk{{\rlap.}^\circ}}
\def\damin  {\wisk{{\rlap.}^\prime}}
\def\dasec  {\wisk{{\rlap.}^{\prime\prime}}}
\def\approxeq{$\sim \over =$}
\def\abouteq{$\sim \over -$}
\def\percm{cm$^{-1}$}
\def\percmsq{cm$^{-2}$}
\def\percmcub{cm$^{-3}$}
\def\perhz{Hz$^{-1}$}
\def\perpc{$\rm pc^{-1}$}
\def\persec{s$^{-1}$}
\def\peryr{yr$^{-1}$}
\def\te{$\rm T_e$}
\def\tenup#1{10$^{#1}$}
\def\to{\wisk{\rightarrow}}
\def\thin{\thinspace}
\def\uk{$\rm \mu K$}
\def\p{\vskip 13pt}

\begin{abstract}
 We measured the dipole of the diffuse $\gamma$-ray background (DGB) identifying a highly significant time-independent signal coincidental with that of the Pierre Auger UHECR.
The DGB dipole is determined from flux maps in narrow energy bands constructed from 13 years of observations by the Large Area Telescope (LAT) of the {\it Fermi} satellite.
The $\gamma$-ray maps were clipped iteratively of sources and foregrounds similar to that done for the cosmic infrared background. The clipped narrow energy band maps were then assembled into one broad energy map out to the given energy starting at $E=2.74$ Gev, where the LAT beam falls below the sky's pixel resolution. 
Next we consider cuts in Galactic latitude and longitude to probe residual foreground
contaminations from the Galactic Plane and Center. 
In the broad energy range $2.74 < E\leq115.5$ GeV the measured dipoles are stable with respect to the various Galactic cuts, consistent with an extragalactic origin. The $\gamma$-ray sky's   
dipole/monopole ratio is much greater  than that expected from the DGB clustering component and the Compton-Getting effect origin with reasonable velocities. At $\simeq (6.5-7)\%$ it is similar to the Pierre Auger UHECRs with $E_{\rm UHECR}\ge 8$ EeV 
pointing to a common origin of the two dipoles. However, the DGB flux associated with the found DGB dipole reaches parity with that of the UHECR around $E_{\rm UHECR}\le 1$ EeV, perhaps arguing for a non-cascading mechanism if the DGB dipole were to come from the higher energy UHECRs.
The signal/noise of the DGB dipole is largest in the $5-30$ GeV range, possibly suggesting the $\gamma$-photons at these
energies are the ones related to cosmic rays.
\end{abstract}

\section{Introduction}

The energy-spectrum of the diffuse gamma-ray background (DGB) is known accurately from {\it Fermi}-LAT measurements: $ F_{\rm DGB}\equiv E^2dN/dE=4\times10^{-7}\left(\frac{E}{\rm Gev}\right)^{-\gamma} {\rm Gev/cm^2/s/sr}$ at $0.1 \lsim E({\rm Gev})\lsim100$ with $\gamma\simeq 0.35$ \citep{Ackermann:2012,Ackermann:2015}. The power of its angular anisotropies have been measured to be flat (white-noise) at $C_\ell/F_{\rm DGB}^2\simeq (8-9)\times 10^{-6}$sr out to $\theta\sim 7^\circ$ ($\ell>50$) \citep{Ackermann:2012a,Ackermann:2018}. The DGB dipole has not yet been measured although it would carry important cosmological information arising from one of the following three origins: 1. The Compton-Getting \citep{Compton:1935} effect from Sun's motion generates an amplified dipole over that of the CMB, assuming the latter is entirely kinematic at $V_{\rm CMB}=370$ km/s \citep{Kogut:1993}.  Since $E^{-2}dN/dE$ is Lorentz invariant \citep{Peebles:1968} the dimensionless kinematic DGB dipole would be ${\cal D}_{\rm DGB}=(4+\gamma)\frac{V}{c}\simeq 0.56\% (V/V_{\rm CMB})$, an amplification which is comparable \citep{Maoz:1994} to that for cosmic rays (CR) \citep{Kachelriess:2006} and the cosmic infrared background \citep{Kashlinsky:2005,Kashlinsky:2022}.  2. The clustering (measured to have white noise angular spectrum) component dipole from the unresolved DGB sources would be $\simeq \sqrt{C_\ell}/F_{\rm DGB}\simeq 0.25\%$ using measurements from \cite{Ackermann:2012,Ackermann:2018} and also here. 3. The Fermi-LAT photons may trace  trace other cosmic structures, such as implied by the UHECR Pierre-Auger dipole of $\simeq (6.5-7)\% F_{\rm DGB}$ \citep{Pierre-Auger-Collaboration:2017,Aab:2020a}.

In proper interpretation the DGB dipole must be described not only with amplitude, but also with its direction. Here, however, fine measurements at high signal/noise $S/N$ are required since the directional uncertainty in the limit of $S/N>1$ is $\Delta \Theta \simeq \sqrt{2}(S/N)^{-1}$rad \citep{Fixsen:2011} and only CMB dipole, probed at $S/N>200$ \citep{Kogut:1993,Fixsen:1994}, currently allows high-precision directional determination.

We present the first measurement of the source-subtracted flux dipole over [3--100]GeV using {\it Fermi} 13 year data. We assemble the UltraClean {\it Fermi} dataset using HEALPix with $N_{\rm side}=128$ \citep{Gorski:2005} from the 13-year all-sky observations at narrow energy bands. After removing known sources, each $E$-map is cleaned of the Ecliptic Plane, the Galactic plane, and remaining Galactic and extragalactic sources, structured foregrounds and noise excursions via a standard iterative procedure used in cosmic infrared background (CIB) work \citep[see e.g. reviews by][]{Kashlinsky:2005,Kashlinsky:2018}. Then possible remaining Galaxy emissions are further removed with progressive Galactic cuts in $(l_{\rm Gal},b_{\rm Gal})$. The clipped narrow $E$-maps were assembled into broad bands of increasing upper energy and the residual map dipoles evaluated for each situation.  We find a robust remaining source-subtracted dipole which is highly statistically significant and  independent of further Galaxy and source removal, consistent with an extragalactic origin. The signal appears stable in time for at least 6 years. The relative amplitude of the dipole is (6--7)\%, which is much higher than could come from the DGB source clustering component and also higher than any kinematic Compton-Getting component, unless the local velocity is $V{>\atop\sim} 3,000$ km/sec. However, the relative dipole amplitude coincides with that of the UHECR dipole from Pierre Auger Observatory \citep{Pierre-Auger-Collaboration:2017,Aab:2020a} which we interpret as indicative of a common origin of the [3--100]GeV photons here and the UHECR.

\section{Fermi-LAT data and processing}

The Fermi Large Area Telescope (LAT) is a pair-conversion telescope covering the nominal $[0.1-1000]$ GeV energy range 
 \citep{Atwood:2009}. It is operated as a sky-survey instrument, having both a wide field of view, $\sim$2 sr and large effective area: $\sim 0.7$m$^2$ for $\sim$1 GeV photons at near normal incidence. The telescope is comprised of a $4 \times 4$ assembly of modules, that each consist of a tracker and calorimeter to record information needed to reconstruct photon direction and energy. The instrument is enclosed by an active anticoincidence system that allows the onboard electronics to reject charged particle events. 

We selected data taken at the start of scientific operations, continuing through the end of mission Cycle 13, specifically, all the weekly photon files 08/31/2008--09/01/2021.  We used the weekly photon files resulting from the latest reprocessing, P8R3\footnote{https://fermi.gsfc.nasa.gov/ssc/data/access/} and the front-plus-back converting events.
To limit contamination from $\gamma$-rays scattered from Earth's atmosphere, a selection cut was applied to remove data taken when the LAT boresight rocked to $>52^\circ$ with respect to the zenith. The selection cuts resulting from this reprocessing significantly reduce the occurrence of cosmic ray induced spurious (i.e., non photon) events, and importantly, remove residual associated anisotropies \citep{Bruel:2018}. It should also be noted that these spurious events can introduce anisotropies in large- or all-sky analyses.  This is due to both non-interacting heavy ions and to cosmic-ray electrons leaking through the ribbons of the Anti-Coincidence Detector, the latter source being responsible for the background anisotropy. 
We also excluded any time interval when the LAT was not in survey mode. To minimize potential contamination from cosmic ray induced events in the detector we selected the ``UltraClean" event selection cut and the corresponding instrument-response function, IRFs. 

The possible effects on the measured DGB dipole by emissions emanating from within the Solar System are demonstrated below to be negligible.


Our analysis benefits from including as many celestial photons as possible but bright sources above our latitude cuts that are in relatively close proximity would likely introduce anisotropies. Thus we exclude emission from the remaining brightest 3,000+ sources from the 4FGL catalog \citep{Abdollahi:2022} sources then employing the clipping procedure as detailed below. 

The $\sim$GeV energy $\gamma$-ray sky at low Galactic latitudes is dominated by diffuse emissions from interactions of cosmic rays with the interstellar medium as well as with the Milky Way radiation fields. While these foreground emissions, which have been studied extensively \citep{Ackermann:2012}, provide a useful diagnostic tool to study the interstellar medium and cosmic ray propagation they are an impediment to large scale extragalactic studies. The Galactic latitude $b_{\rm Gal}$, and longitude $l_{\rm Gal}$, cuts employed in our analysis partially remove the effects of these structured foregrounds on our analysis. We assess further effects by making incremental cuts at $|b_{\rm Gal}| > 20^\circ, 30^\circ, 45^\circ$ and $30^\circ,45^\circ \leq l_{\rm Gal}\leq 330^\circ, 315^\circ$, removing the Ecliptic Plane and recomputing the dipole. 

We apply the clipping procedure well known and commonly used in the source-subtracted CIB studies since the COBE/DIRBE, later refined for and further tested in the various 2MASS, {\it Spitzer} and {\it Euclid} work \citep[see reviews][]{Kashlinsky:2005,Kashlinsky:2018}. The method isolates iteratively the pixels with photons exceeding a given threshold of $N_{\rm phot}(l_{\rm Gal}, b_{\rm Gal})\geq \langle N_{\rm phot}\rangle + N_{\rm cut} \sigma(N_{\rm phot})$, removing here the entire beam at 95\% c.l. around the identified in the given iteration ``sources" and proceeds until no more such excursions are found at the given $N_{\rm cut}$. Typically up to 10 iterations were needed to converge and the results are insensitive at $N_{\rm cut}\leq 4$. Of course, the method should be applied to as narrow bands as possible, especially for the $\gamma$-ray sky where adjacent energies often trace different sources.
Figs \ref{fig:fig1a_appx},\ref{fig:fig1b_appx} show the commonality of the remaining photons for the unclipped and clipped maps demonstrating the efficiency and necessity of clipping. After clipping, most remaining pixels starting already around (3--5) Gev trace different sources at two adjacent bands, and most of the sky becomes progressively darker (no photons in each pixel) at narrow energy bands, as expected. Next we combine different bands into one broader band map and then evaluate the dipoles and the mean remaining flux.

\begin{figure}[h!]
\includegraphics[width=6in]{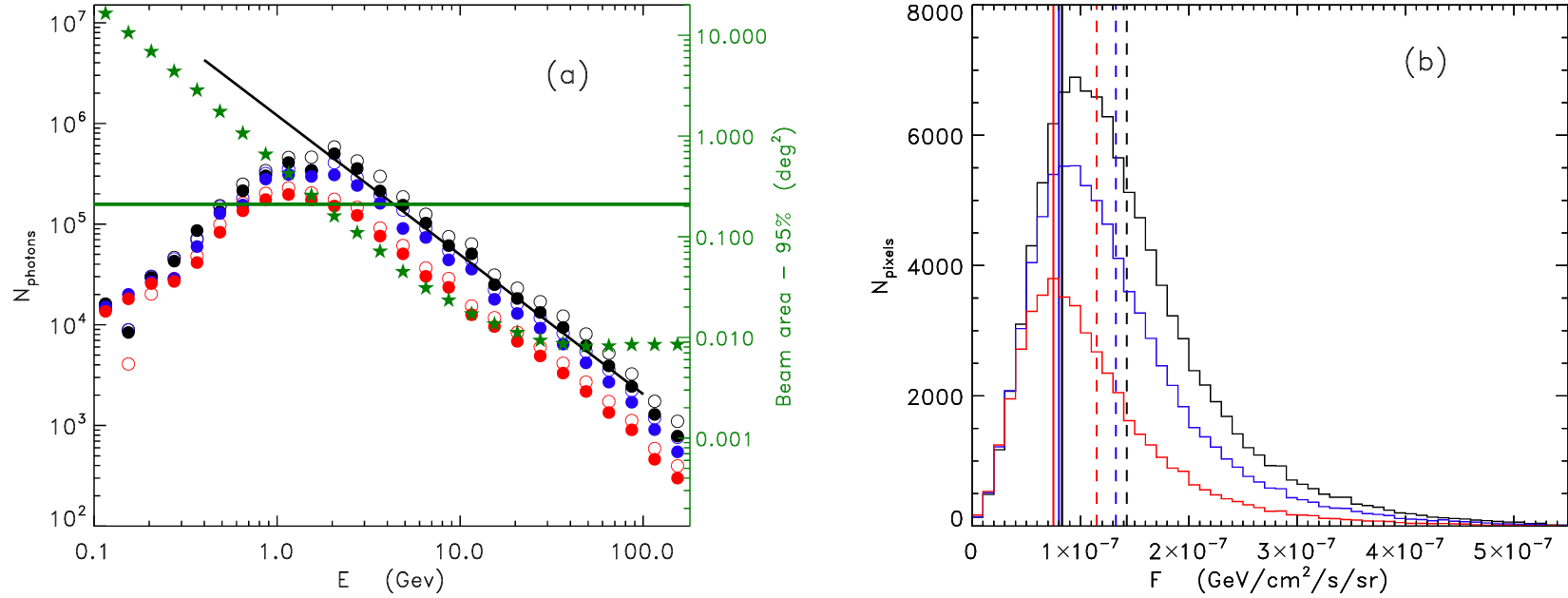}
\caption{(a) Total number of photons in each narrow band is marked with filled circles. Black, blue, red circles correspond to $|b_{\rm Gal}|\geq 20^\circ,30^\circ,45^\circ$ where filled(open) circles to $0\leq l_{\rm Gal}\leq 360^\circ$($30^\circ\leq l_{\rm Gal}\leq 330^\circ$). The LAT beam 
 is shown with green asterisks for the values displayed on the right vertical axis. The horizontal thick green solid line marks the pixel area for $N_{\rm side}=128$. The slope of the DGB is marked with black solid line. (b) Flux histograms for $2.74 \leq E({\rm Gev})\leq 115$ and $30^\circ \leq l_{\rm Gal} \leq 330^\circ$ with black/blue/red corresponding to  $|b_{\rm Gal}| \geq 20^\circ,30^\circ,45^\circ$. Dashed and solid vertical lines of 3 colors mark the mean flux and flux dispersion of each map. 
}
\label{fig:fig1}
\end{figure}
Fig.\ref{fig:fig1}a shows the number of photons left in each band after the procedures. The horizontal green line in Fig.\ref{fig:fig1}a shows the pixel area (right vertical axis) and the green asterisks mark the 95\% LAT beam area which falls below our pixel resolution at $E\gsim 3$GeV. The remaining after clipping $\gamma$-ray sources are unresolved at this resolution. 

\begin{figure}[h!]
\includegraphics[width=6.5in]{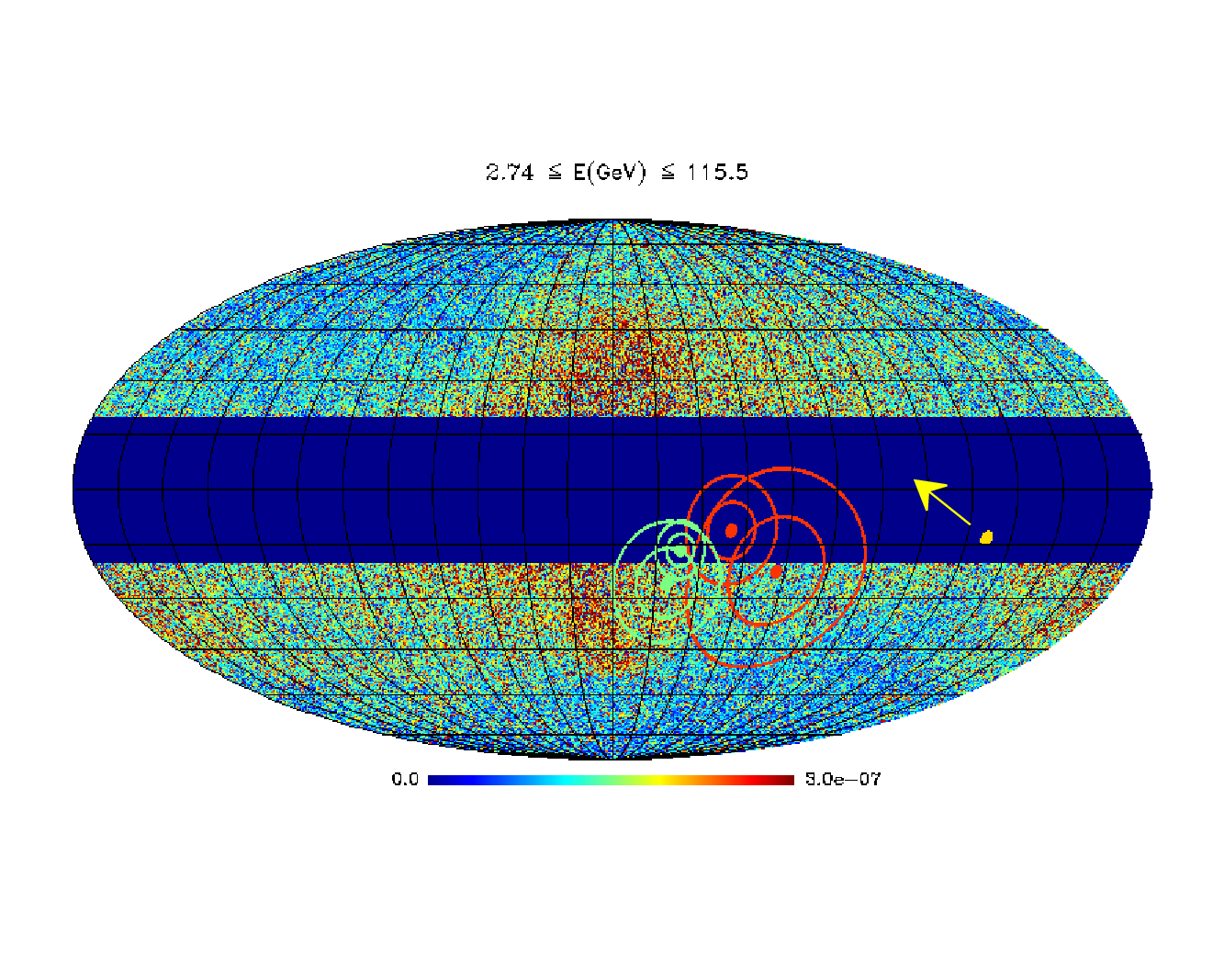}
\caption{\small The source-subtracted LAT maps coadded from 2.74 to 115 GeV. The displayed sky is masked for $|b_{\rm Gal}|\leq20^\circ$. The coordinate grid is marked every 15$^\circ$. 
The DGB dipole directions are plotted for the $45^\circ\leq l_{\rm Gal}\leq  315^\circ$,
$|b_{\rm Gal}| \geq 45^\circ$ with red colors and the $30^\circ\leq l_{\rm Gal}\leq 330^\circ$, $|b_{\rm Gal}|\geq 30^\circ$ with green
colors. The two contours in each case correspond to $1\sigma$ and $2\sigma$ (68\% and 95\%) dipole uncertainties. The size of each contour decreases as it should when larger sky area is considered.
The smaller/narrower contours indicate the directions and uncertainties of the 13 yr assembled map shown here. The larger/broader contours show the same once the $(A-B)/2$ dipoles have been subtracted. 
The yellow arrow indicates the tentative position of the Pierre Auger dipole prior 
to the deflection by the modeled Galactic magnetic field. The Pierre Auger UHECR 
dipole is marked with the yellow spot and its 68\% directional uncertainty is 
$\sim 15^\circ$. The map with the Eclipic plane masked out with $|\beta_{\rm Ecl}|\leq 5^\circ$ is shown in Fig. \ref{fig:fig3_appx}.
} 
\label{fig:fig2}
\end{figure}
We therefore restrict our dipole DGB analysis to $E> 2.74$ GeV. The clipped maps at narrow energy bands are then coadded up to $E=115$ GeV. Fig.\ref{fig:fig1}b shows the consistency of the histograms of the assembled source-subtracted maps for various $|b_{\rm Gal}|$-cuts and $30^\circ\leq l_{\rm Gal}\leq 330^\circ$.
The resultant clipped map assembled over $2.74\leq E({\rm GeV})\leq 115$ is shown in Fig.\ref{fig:fig2}; the symbols are explained below.

\section{Dipole results}

Computations of the multipole expansion of source-subtracted, band-added maps were done with standard HEALPix routines:  dipole with REMOVE$\_$DIPOLE and power spectrum with ANAFAST developed with singular-value-decomposition \citep{Gorski:2005}.

\begin{figure}[h!]
\includegraphics[width=5in]{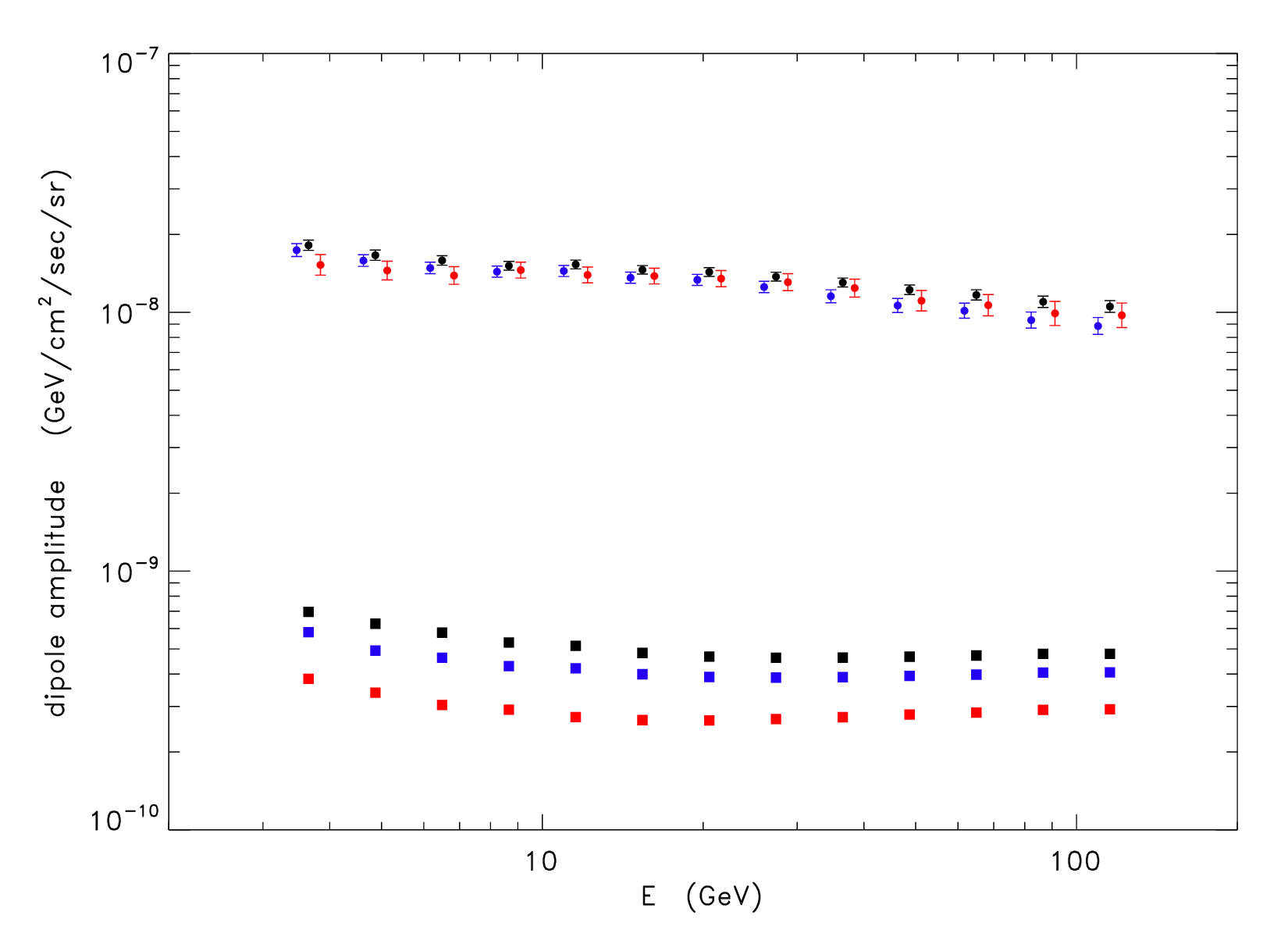}
\caption{\small  Black, blue, red colors mark $|b_{\rm Gal}|\geq 20^\circ, 30^\circ, 45^\circ$ and $30^\circ \le l_{\rm Gal} \le 330^\circ$. The points are slightly shifted at each $E$ for clearer display and the configuration of $45^\circ \le l_{\rm Gal} \le 315^\circ$ is not shown to avoid clutter (it is shown in the next figure).
The square-root of the clustering mean power over $\ell \geq 20$ from Fig. \ref{fig:fig3} are shown with squares. The dipoles in the source-subtracted 13-year maps are shown with filled circles with the 68\% uncertainties.}
\label{fig:fig3}
\end{figure}

The dipole from the clipped maps is shown with circles  in Fig.\ref{fig:fig3}.
The dashed lines of three colors in Fig. \ref{fig:fig1}b show the mean flux in the clipped maps of 3 values of $b_{\rm cut}$ and the solid lines of the same color show the flux dispersion of each map. The flux dispersion is $\sigma_F \sim 0.8\times 10^{-7}$ GeV/cm$^2$/s/sr which introduces random dipole uncertainty in each of the 3 components over a map of $N_{\rm pix}$ pixels of $\sim \sigma_F/\sqrt{N_{\rm pix}}$. For the dipole amplitude the uncertainty will be $\sqrt{3}\sigma_F \sim 5\times10^{-10}\sqrt{N_{\rm pix}/10^5}$ GeV/cm$^2$/s/sr which shows that the dipoles found in Fig.\ref{fig:fig3} are statistically significant. 
A more accurate estimate of the statistical error can be done by bootstrap.
We generated one thousand simulated maps by assigning to each point on the unmasked regions of
the sky a value extracted at random from the flux map of 13 years of data. All data points
were given the same probability. Following bootstrasp standard practice,
the selected values were not removed. Once a random map $j$ was generated its dipole
$\vec{d}^{\, j}$ was computed using the HEALPix remove\_dipole routine.
The mean of each dipole component is zero and its dispersion is a measure of
the statistical error. The ratio of the signal to its statistical uncertainty
is largest in the energy range $\sim(5-30)$GeV.

Next, we computed the statistical uncertainty in the dipole direction using the errors
$\sigma_i$ by generating $k=10^4$ gaussian distributed vectors $\vec{d}^k$
about the dipole components $D_i$ measured from the 13-year flux map, 
$d_i^k=D_i+r_i^k\sigma_i$ with $r_i^k$ gaussian random numbers with zero 
mean and unit variance. These bootstrap error estimates do not probe time-variable 
component of the dipole that might be present in the data from instrument noise,
systematics or genuine 
$\gamma$-ray variability. 

For estimating the time-variations and noise in the maps 
we assembled 2 time-separated subsets of data: 
A for years 1-6 and B for years 7-12 with the odd year 13 left out 
temporarily for this task. Each of the two 6-yr time intervals are longer than 
the typical variability timescales of $\gamma$-ray sources. 
The masks from the clipping of sources and Galaxy for the full 13-year map in
$(b_{\rm Gal},l_{\rm Gal})$ are applied to each subset and the narrow E-bands are 
then coadded to produce the 13 wide band maps going from $E_0=2.74$GeV
to the given $E$ up to $E=115$GeV. Then, similar to the by-now standard
CMB \citep{Smoot:1992,Bennett:1996} and CIB \citep{Arendt:2010,Kashlinsky:2012a} 
noise power processing, the noise properties are
computed from the $(A-B)/2$ 
maps, masked in the same way as the combined 13 year data. 
The resultant noise map is shown in Fig. \ref{fig:fig5_appx} and the dipole
amplitudes of the 13yrs and $(A-B)/2$ and dipole powers are compared in 
Fig. \ref{fig:fig11_appx}. Fig. \ref{fig:fig6_appx}, left displays the histograms in the time-differenced maps with the Ecliptic Plane masked out showing the standard deviations  using 12yr data of $\sigma_{F,(A-B)/2}\sim 1.2\times10^{-7}$ GeV/cm$^2$/s/sr. The right panel plots the correlation coefficient between the $(A-B)$ and the final 13 yr maps, and demonstrates negligible to null correlations implying that the two maps are independent with the dipole power from the $(A-B)/2$ contributing in quadrature to the signal shown in Fig. \ref{fig:fig3}. 

Regarding the possibility of systematics, our analysis is not dependent on a precision spectral determination or on an absolute flux both of which are susceptible to uncertainties in the instrument response. Furthermore, we are not subtracting a Galactic foreground model which may introduce potential systematics.  Finally, our error analysis based on the $A-B$ time differences argue against the presence of time dependencies systematics which could arise from systematic effects in the exposure maps. 


We evaluated the power spectrum of the {\it dipole- and monopole-removed} source-subtracted 13-year maps and find it is approximately white out to the computed $\ell=500$ in agreement with the earlier analyses \citep{Ackermann:2012,Ackermann:2018}, except about 20-25\% lower because of our (more efficient) source subtraction. The clustering dipole contribution, from averaging the (white) power at $20 < \ell  <500$ to reduce the effects of the cosmic variance at low $\ell$, is shown with squares in Fig. \ref{fig:fig3}. For the clustering independent of the uncovered dipole, the latter would add in quadrature contributing negligibly  in the final dipole balance compared to the found dipole.

The angular uncertainty on the measured $\gamma$-ray dipole is
shown in Fig. \ref{fig:fig2} for two configurations
$|b_{\rm Gal}|=30^\circ$, $30^\circ\le l_{\rm Gal}\le 330^\circ$ (green contours),
and $|b_{\rm Gal}|=45^\circ$, $45^\circ\le l_{\rm Gal}\le 315^\circ$ (red contours).
To estimate the directional uncertainty 
we likewise generated $k=10^4$ random dipoles $\vec{d}^k$ with the
uncertainties derived from bootstrap. The angular separations between
the measured and simulated dipoles sorted in increasing order and
and the 1-,2-$\sigma$ confidence intervals 
were defined as the regions that encloses 68\% and 95\% of all angular 
separations. The uncertainties are marked by the two sets (red and green) 
of small contours. This dipole determination includes diffuse and 
time-variable contributions. 
To estimate the direction and its uncertainty of the diffuse $\gamma$-ray 
component alone, we subtracted the dipole vector $\vec{D}_{(A-B)/2}$ from the 
13yrs dipole vector $\vec{D}_{\rm 13yrs}$ and added the errors in quadrature. We
repeated the procedure as before and the results are shown by the wider
red and green contours. Although the 
time varying $\gamma$-ray component on the direction has some effect,
in Appendix we show that its effect on 
the dipole amplitude is negligible. The Pierre Auger dipole is shown   
as a yellow point and the arrow indicates its possible location
prior to the deflection produced by a model of the Galactic magnetic field.

\section{Discussion}

\begin{figure}[h!]
\includegraphics[width=5.5in]{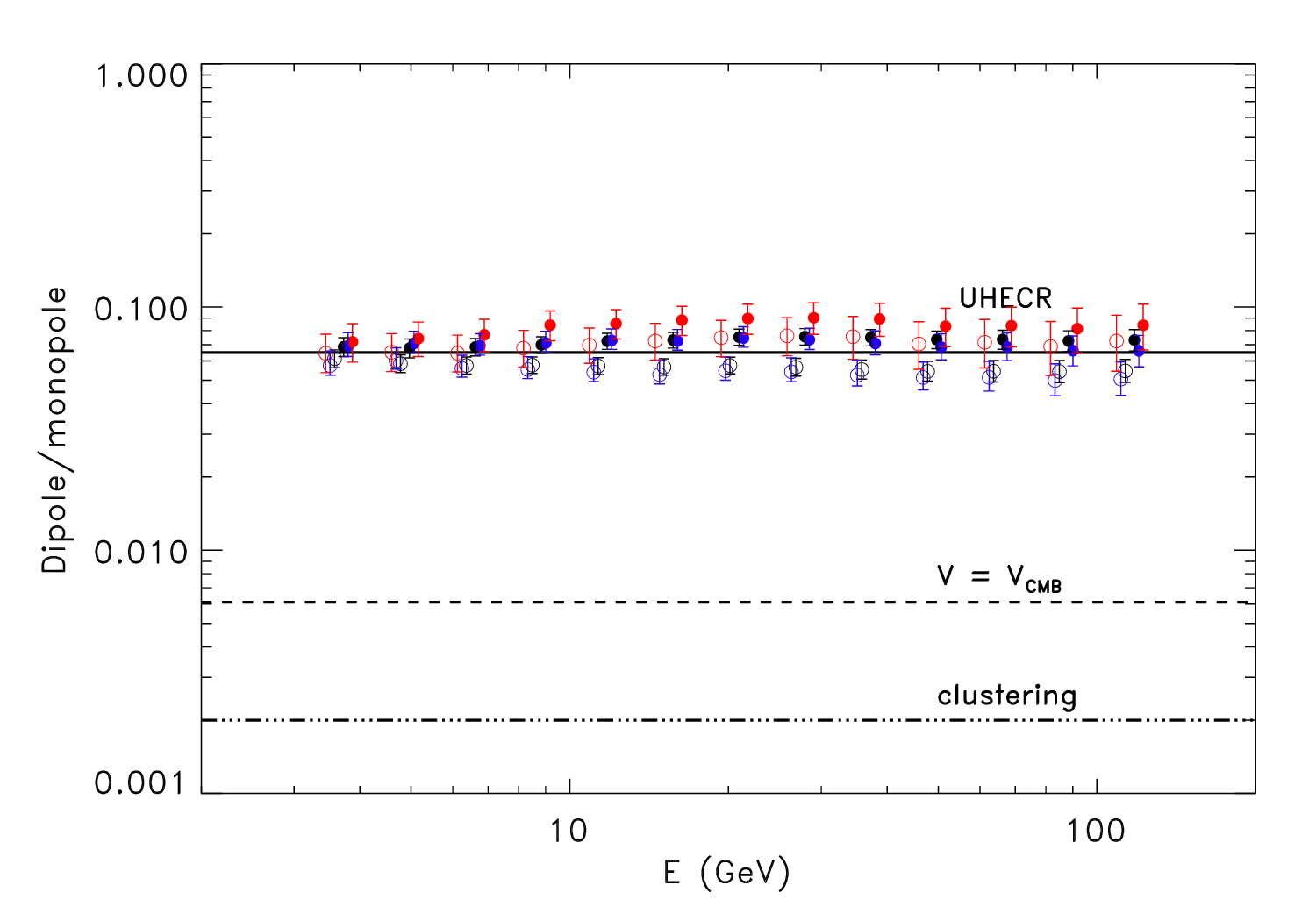}
\caption{\small 
Dimensionless dipole in the same notations as in Fig. \ref{fig:fig3}. In addition to the numbers in Fig. \ref{fig:fig3}, red/blue/black open circles denote the configurations with $45^\circ\leq l_{\rm Gal}\leq 315^\circ$ and $2.74 < E({\rm GeV})<115$ slightly shifted in $E$ for clear viewing (this is not shown in the earlier figure to avoid overcrowding - all the numbers are consistent with what is shown here). The dimensionless dipoles from cosmic rays (UHECR) at 6.5\%, Compton-Getting amplified CMB dipole and clustering component are marked with horizontal lines. Removing the Ecliptic Plane, as discussed in the Appendix, further reduces the scatter - see Fig. \ref{fig:fig4_appx}. The statistical uncertainties are shown at 95\% c.l.
}
\label{fig:fig4}
\end{figure}
Fig. \ref{fig:fig4} summarizes our findings in terms of the dimensionless dipole amplitudes. The identified dipole has dimensionless amplitude $\simeq (6.5-7)\%$ out to $E\sim 100$ GeV and can be either of Galactic origin, or extragalactic from one of the 3 origins mentioned in the Introduction. It is unlikely to come from the Galaxy as its properties show no dependence on the Galactic cuts, $|b_{\rm Gal}|, l_{\rm Gal}$, outside of the Galactic Center for $E\gsim 3$GeV. 
Foregrounds are removed efficiently here and the DGB is reproduced for the remaining sky as Fig.\ref{fig:fig2} showed. 

The dipole directions are consistent for
all Galaxy cuts and $E$-configurations with the 68\% c.l. errors of $\sim$10$^\circ$--20$^\circ$. The rms directional uncertainty is close to $\Delta\Theta\sim \sqrt{2}(S/N)^{-1}$ radian \citep{Fixsen:2011}.
For data in the region $30^\circ \le l_{\rm Gal} \le 330^\circ$
and cuts in $|b_{\rm Gal}|=20^\circ,\; 30^\circ,\; 45^\circ$, the dipoles are located in the range $(l_{\rm Gal},b_{\rm Gal})=(-30^\circ\rightarrow \sim 0)$; for the data in the 
region $45^\circ\le l\le 315^\circ$ and for the same cuts in $|b_{\rm Gal}|$ the dipoles
are at Galactic $(l,b)=(270^\circ\rightarrow 300^\circ,-40^\circ\rightarrow 0)$. 
In Fig.~2 we show the direction and its 68\% and 95\%
angular dispersion 
for several configurations. 
We note that the direction comparison is further hampered by the deflection
of cosmic rays by the Galactic magnetic field, shown by the 
(model dependent) location of the UHECR dipole prior to this effect.

We tested for Solar System effects, primarily from the Sun and Moon, by removing the Ecliptic Plane and find no noticeable changes in the dipole as discussed in Appendix. The distribution of Solar-system emissions  are  expected to be sharply distributed about the ecliptic \citep{Johannesson:2013}. We considered the possibility of $\gamma$-ray emission emanating from the Solar system by masking the $10^\circ$ band of the Ecliptic Plane  and find the same signal within the errors, indicating negligible contribution from there demonstrating at most a negligible dipole contribution to the measured signal.

The expected dipole component from clustering, $d_{\rm clustering}$, is shown in Fig.\ref{fig:fig3} to be on average $\simeq 0.2\%$ from using the power spectra computed here which is consistent with \cite{Ackermann:2012,Ackermann:2018}; if the found dipole is independent of clustering  $d_{\rm clustering}$ is added in quadrature making this contribution totally negligible. Cosmic variance affects the measured clustering dipole per probability distribution $p(d_{\rm clustering})=\sqrt{\frac{\pi}{2C_1}}\frac{d_{\rm clustering}^2}{C_1}\exp\left(-\frac{d_{\rm clustering}^2}{2C_1}\right)$  \citep{Kaiser:1983,Abbott:1984}; hence at 95, 99\% c.l. $d_{\rm clustering}<4, 5.7 \sqrt{C_1}$, still well below the found dipole. This makes clustering an unlikely origin for the measured dipole. The fact that the measured diffuse emission power spectrum agrees well with the other measurements and shows no drastic dependence on the Galaxy cuts further indicates that the diffuse emission, and likely its dipole, does not have critical Galactic components. 

If the dipole originates from the Compton-Getting effect due to our kinematic motion the implied velocity would be $\gsim 3,000$ km/sec, an order of magnitude larger than the CMB dipole equivalent velocity marked with the dashed line in Fig. \ref{fig:fig4} and significantly larger than the effective velocities implied in other dipole measurements \citep{Kashlinsky:2008,Kashlinsky:2010,Kashlinsky:2011,Singal:2011,Atrio-Barandela:2015,Secrest:2022}.  Thus the Compton-Getting amplified kinematic origin of the measured DGB dipole also appears to us implausible.

At face-value there appears a commonality with the UHECR dipole from Pierre Auger Observatory \citep{Pierre-Auger-Collaboration:2017}. The dimensionless amplitude of the UHECR dipole at $E\gsim 3$ EeV is shown in Fig. \ref{fig:fig4} with the solid line at 6.5\%. It coincides well with the found DGB dipole perhaps suggesting a common origin. 
The source of the UHECRs in general is still unknown. Active galactic nuclei (AGN) have long been considered as potential sites for production of high-energy cosmic rays including possibly UHECRs \citep{Fang:2018,Murase:2022}. Also, there has been similar speculation regarding gamma-ray bursts (GRBs) and they cannot currently be ruled out as sources of prolific UHECR emission \citep{Murase:2022a}. However, both of these populations are well known to be highly isotropic. So, some as yet unidentified subpopulation would be required to produce the dipole anisotropy documented by the Pierre Auger experiment if AGNs or GRBs are its source.  
Very generally a common origin of UHECR and the [3--100] GeV DGB may be due to pionic photons \citep{Halzen:2002} from photomesonic production \citep{Stecker:1999} arising in several mechanisms.  Possible venues may come from UHECRs above the GZK knee around $\sim10^{20}$ eV \citep{greisen:1966,Zatsepin:1966} with pions arising from the $\Delta$-resonance: $p+\gamma_{\rm CMB}\rightarrow \Delta^+ \rightarrow \pi^0+p$, or more speculatively from proton decay \citep{Sakharov:1967} $p\rightarrow e^+ + \pi^0$, $p\rightarrow \mu^++\pi^0$ \citep{Tanabashi:2018}.  Subsequently, $\pi^0\rightarrow 2\gamma$ and on average the pionic photons would carry $\sim 1/2$ of the parent pion's energy \citep{Meszaros:2014,Halzen:2019,Globus:2023}. GZK horizon of $\lsim 100$ Mpc would then apply to the sources in the latter case \citep{Stecker:1968,Ding:2021} whose precise value depends on the CR composition \citep{Ahlers:2011}. 
The pions trigger electromagnetic cascades, transferring energy from the CRs to the GeV--TeV $\gamma$-ray photons \citep{Fornasa:2015}, but their energy flux would have to be up to 50\% of the CR flux \citep{Kalashev:2009}. 
However, the dipole here corresponds to the [3--100] flux of $\sim (1-1.5)\times 10^{-7}$ GeV/cm$^2$/s/sr which is reached for UHECRs around $E_{\rm UHECR}\lsim 1$ EeV \citep{Meszaros:2014,Fang:2018,Aab:2020}, so it is not clear whether a cascading mechanism can produce the $\gamma$-ray signal out of the UHECR protons. More likely this points to a common origin between the cosmic rays and the high-$E$ photons observed here by {\it Fermi}-LAT. See possible specific models in e.g. \citep{Waxman:1995,Fang:2018,Ding:2021}. In this case the distance to the sources may not exceed the horizon for high-$E$ $\gamma$-rays \citep{Nikishov:1961,Stecker:1968,Fazio:1970} from CIB and optical background \citep{Kashlinsky:2005b,Helgason:2012a,Fermi-LAT-Collaboration:2018}. These sources lie at smaller distances than the cluster sample used in the dark flow dipolar probe \citep{Kashlinsky:2008,Kashlinsky:2010}, or the source sample of the WISE dipole probe \citep{Secrest:2022} and cannot affect those analyses.


However, as a result of magnetic deflection, using UHECR anisotropy information to constrain the spatial distribution or the nature of the comic-ray sources is not plausible given the limitations inherent in galactic and extragalactic magnetic field models \citep{Allard:2022}. Indeed, we cannot unambiguously claim an association between the UHECR and gamma-ray dipoles, although the coincidence in amplitude and nominal orientation is intriguing. We would note that though that in the case of a common source the dipole nature of the UHCER flux is likely to be preserved despite magnetic deflection of individual particles.
 
In addition to the issue of magnetic deflection it must be considered that the sources of gamma-rays and UHECRs maybe not be strictly in common. For example, it is apparently the case that some subset of gamma-ray emitting blazar AGN are prolific sources of high-energy neutrinos while
some are not; or perhaps the production sites within the latter are opaque to gamma-ray emission; see e.g. \citep{Plavin:2023}. Additionally, there are apparently non-blazar sources of high-energy neutrinos, e.g., NGC 1068 and perhaps the Milky Way itself. The Ice-Cube neutrinos may be unrelated to the Auger UHECRs, but by analogy the underlying UHCER sources may comprise a similarly non-homogeneous population.  
\section*{Appendix}
\subsection*{Map clipping}

\begin{figure}[h!]
\hspace{-0.75cm}\includegraphics[width=6.75in]{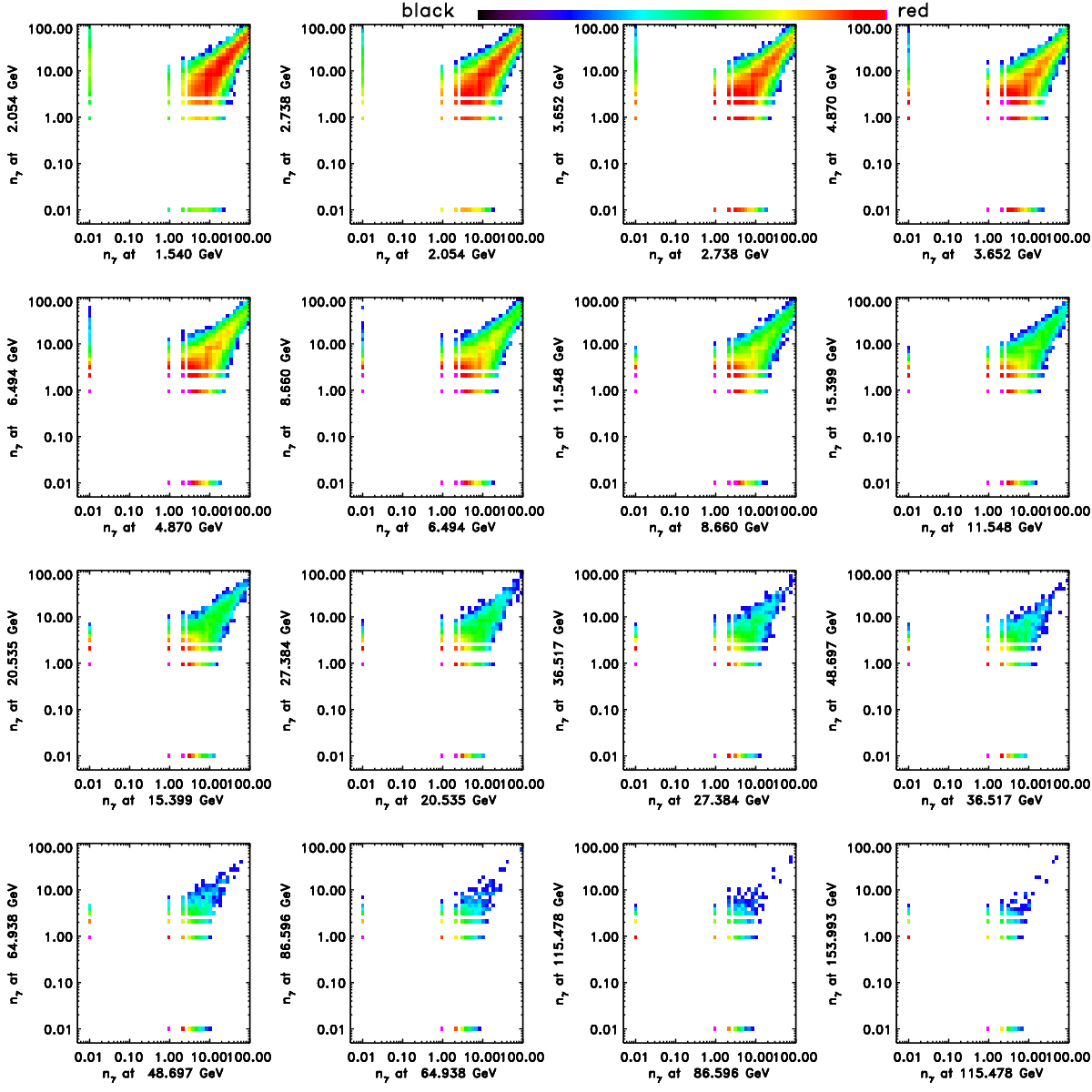}
\caption{\small Number of photons per given pixel common in the two adjacent maps is shown with density plots for unclipped maps. The pixels where there are zero photons in one of the 2 bands are shown at $n_\gamma=0.01$.  The shown colors span 0 to 5,500 common pixels logarithmically.}
\label{fig:fig1a_appx}
\end{figure}
\begin{figure}[h!]
\includegraphics[width=6.75in]{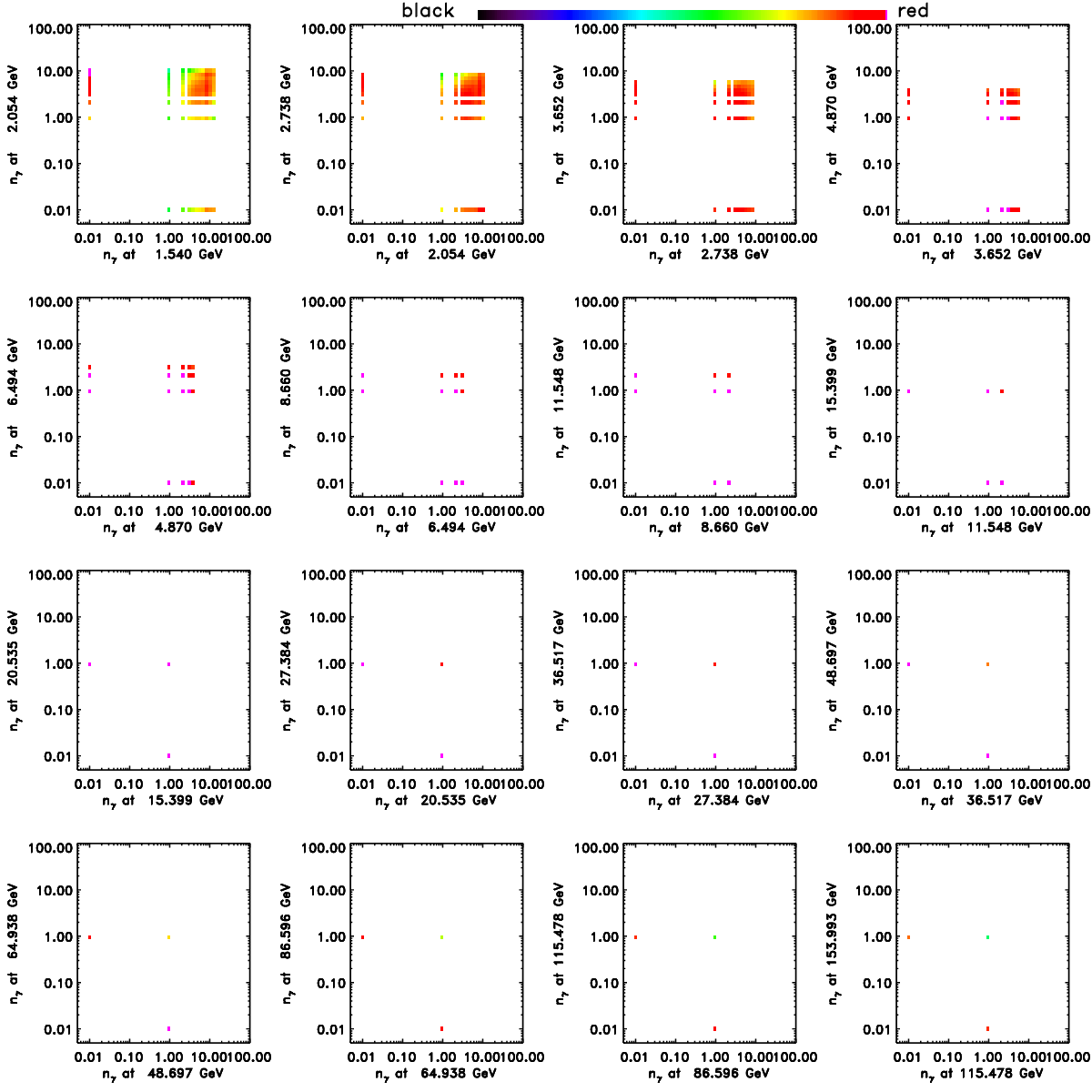}
\caption{\small Same as Fig. \ref{fig:fig1a_appx} but for clipped maps. }
\label{fig:fig1b_appx}
\end{figure}
Fig. \ref{fig:fig1a_appx} shows the commonality of photons in unclipped maps  at the given $E$ between each 2 adjacent bands starting at 1.15 Gev for the energy range at which the DGB dipole will be assessed. Magenta colors correspond to most photons with blue to the fewest. The figure shows that as one reaches higher $E$ there are fewer and fewer common pixels even among the two most adjacent bands.  Fig. \ref{fig:fig1b_appx} shows the same after clipping. 
\clearpage
\subsection*{Testing for null systematics from the plane of the Ecliptic}

In principle, the Sun produces diffuse $\gamma$-rays from the inverse Compton scattering on its radiation field and the Moon
produces some diffuse gamma-ray component from cosmic-ray interactions. Once averaged over the year, the Sun and the Moon may contribute some
anisotropic diffuse components contributions which we test here.

\begin{figure}[h!]
\includegraphics[width=4in,angle=90]{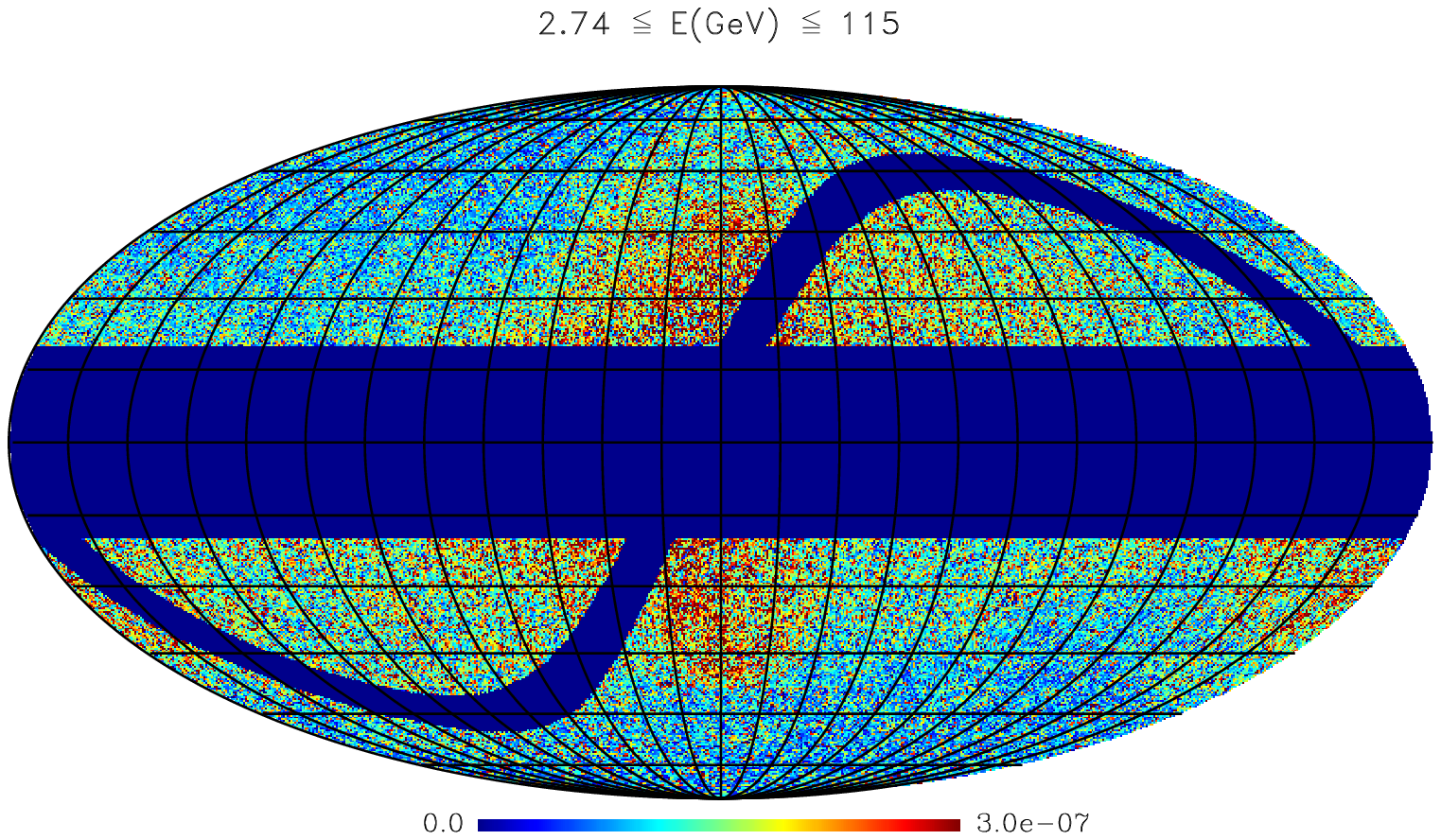}
\caption{\small Map as in Fig. \ref{fig:fig4} without masking the Ecliptic plane.}
\label{fig:fig3_appx}
\end{figure}
From, e.g., \cite{Ackermann:2015} it is evident that the solar system flux can be at a level of up to $\sim 5\%$ of the isotropic background level. However, this emission, which is due to cosmic ray interactions with the Solar disk and the Moon as well as the inverse-Compton radiation involving the Solar radiation field, is concentrated along the Ecliptic Plane; see e.g.  Fig. 2,  \cite{Johannesson:2013}. We also generated emission templates as described therein and examined the latitude profiles. Examination of the resulting latitude profiles suggest that a $10^\circ$ Ecliptic latitude cut should lead to eliminating $>90\%$ of the Solar System emissions. 
In any case, the Ecliptic latitude cut did not lead to a discernible difference in our dipole determination as noted in the main part.

Fig. \ref{fig:fig3_appx} shows the same map as used in the main part, Fig. \ref{fig:fig3}, demonstrating no obvious contributions from emissions within the Ecliptic Plane.

\begin{figure}[h!]
\includegraphics[width=5in]{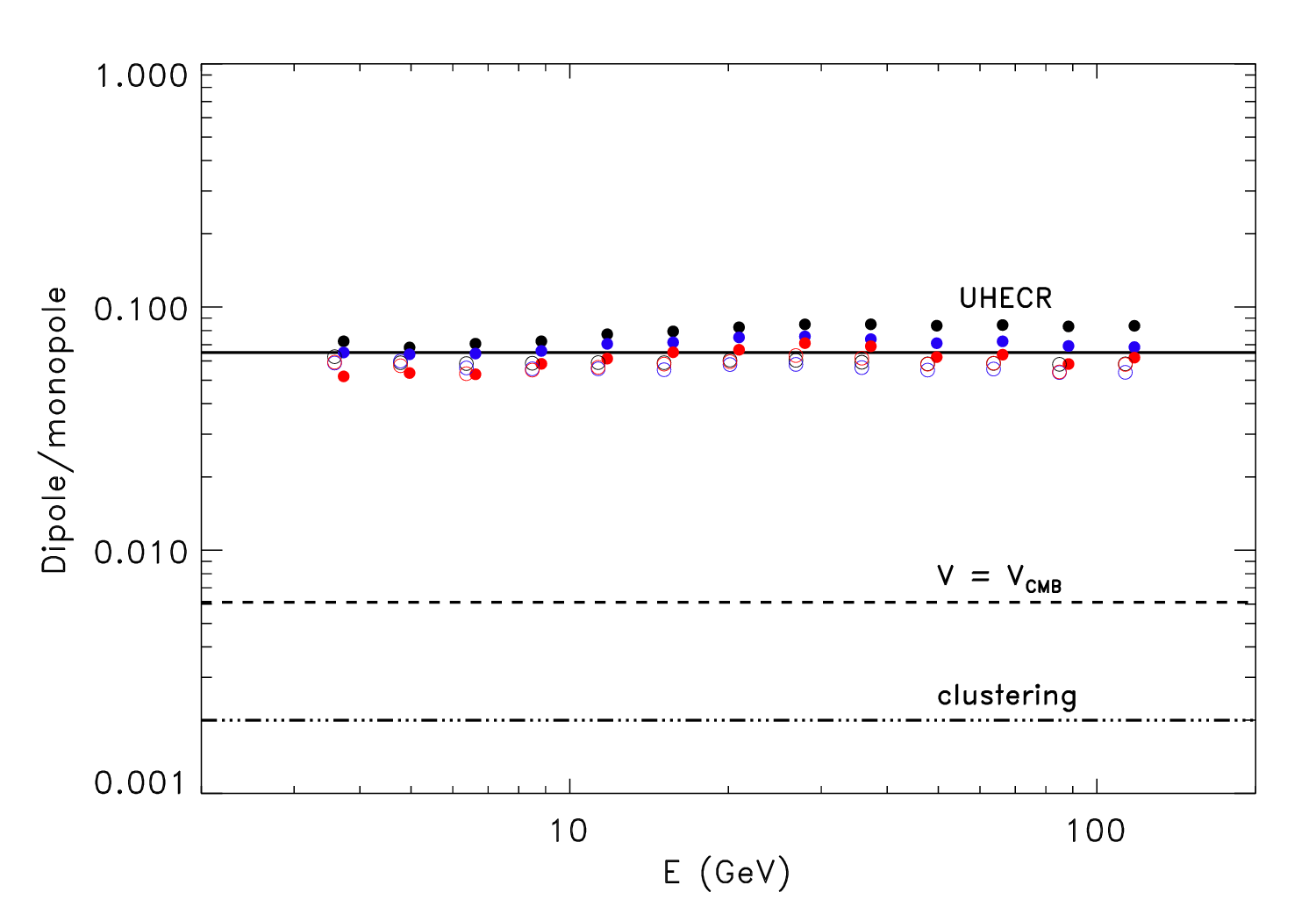}
\caption{\small Dimensionless dipole in the same notations as in Fig. \ref{fig:fig4} without masking the Ecliptic plane. The error bars are barely distinguishable from those in Fig. \ref{fig:fig4} and are not shown to avoid clutter.}
\label{fig:fig4_appx}
\end{figure}

Fig. \ref{fig:fig4_appx} shows the dipole/monopole ratio, as in Fig. \ref{fig:fig4}, when emissions from the Ecliptic Plane are not masked out. The two sets of numbers are essentially identical within the errors demonstrating no substantial contributions to the identified  DGB dipole from the Solar System and Moon $\gamma$-ray emissions.

\subsection*{Testing for time-variability with time differenced maps}

\begin{figure}[h!]
\includegraphics[width=4.in,angle=90]{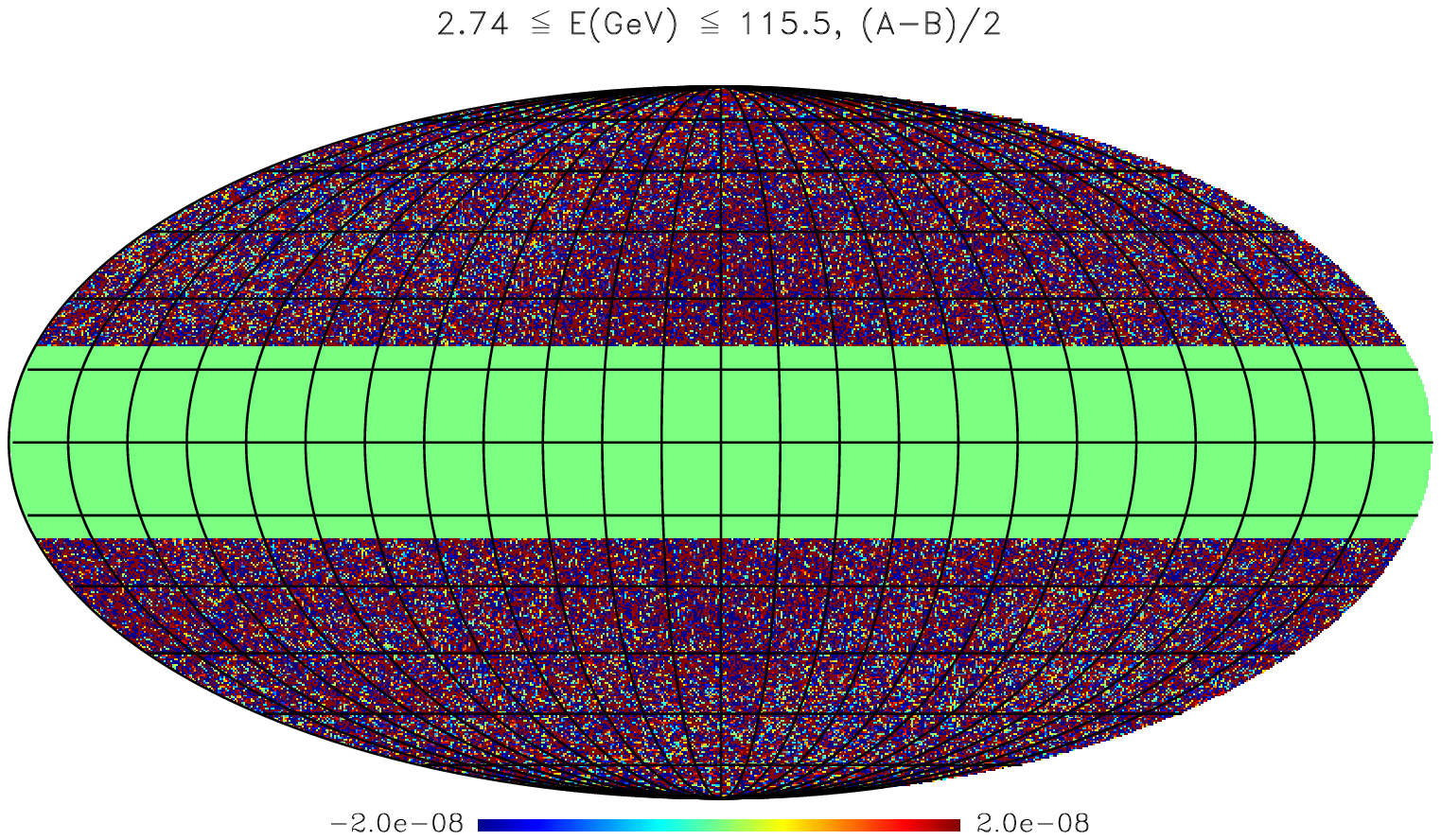}
\caption{\small  The $(A-B)/2$ map from 2.74 to 115 Gev is shown with the overall 13 year clipping mask applied and Galaxy removed for $|b_{\rm Gal}|\leq 20^\circ$.} 
\label{fig:fig5_appx}
\end{figure}
For estimating the time-variations and noise in the maps we assembled 2 time-separated subsets of data. Subset $A$ for years 1--6 and subset $B$ for years 7--12 with the odd year 13 left out temporarily in this task. The two 6-yr time intervals are each longer than the variability timescales of $\gamma$-ray sources. The masks from the clipping of sources and Galaxy for the full 13-year map in $(b_{\rm Gal},l_{\rm Gal})$ are applied to each subset and the narrow $E$-bands are then coadded to produce the 13 band maps going from $E_0=2.74$ GeV to the given $E$ up to the final $E=115$ GeV. 

Fig. \ref{fig:fig5_appx} shows the resultant time-differenced map, $(A-B)/2$, with the Galactic Plane cut out at $|b_{\rm Gal}|= 20^\circ$.

\begin{figure}[h!]
\hspace{-0.5cm}\includegraphics[width=6.5in]{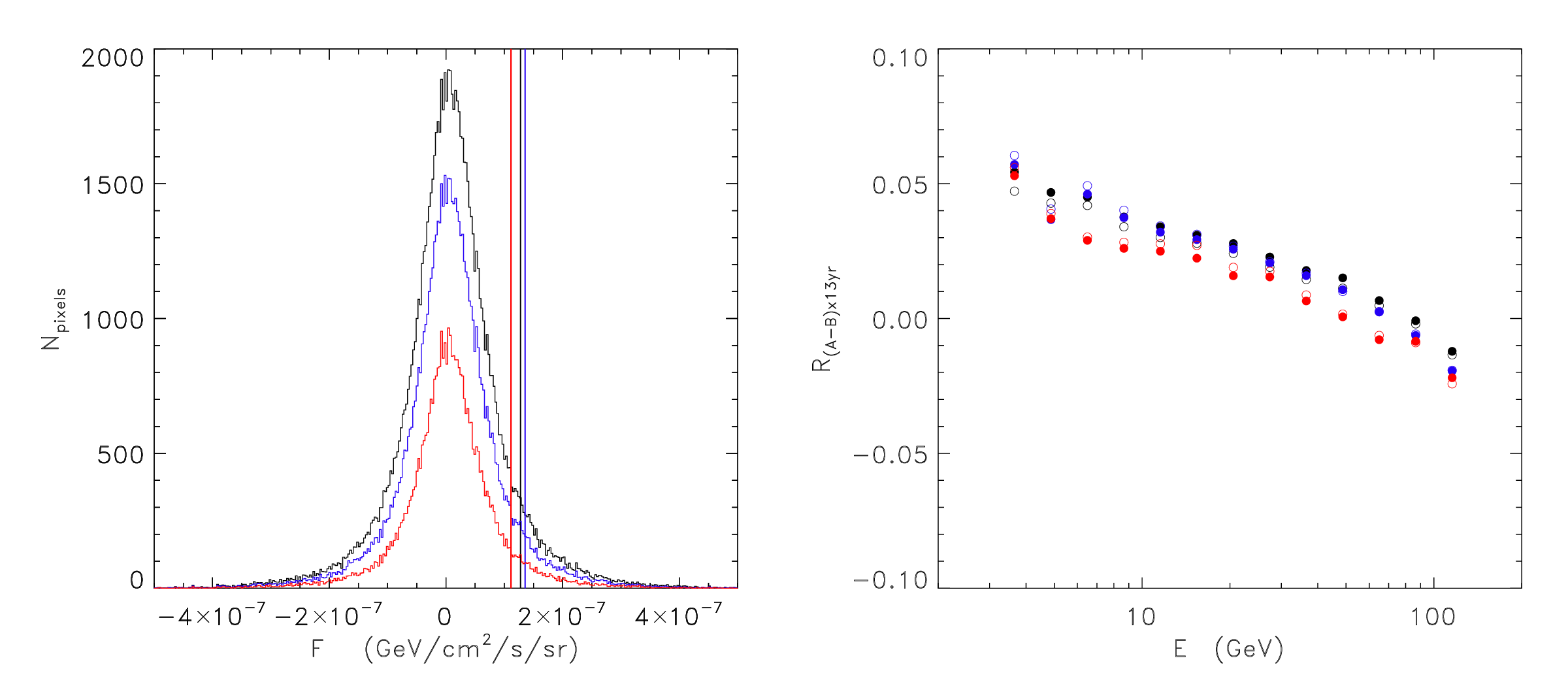}
\caption{\small (Left): Histograms of fluxes after the 13-yr clipping mask has been applied to the maps shown in Fig. \ref{fig:fig5_appx}. The standard deviations are shown with vertical solid lines corresponding to $\sigma_{F,(A-B)/2}\sim 1.2\times10^{-7}$ GeV/cm$^2$/s/sr. The masking from clipping has been applied and the sky has been kept at $30^\circ\leq l_{\rm Gal}\leq 330^\circ$ and $|b_{\rm Gal}|\geq 20^\circ, 30^\circ,45^\circ$ for the black, blue, red colors. (Right) The correlation coefficient, $R$, between the $(A-B)$ and the final 13 yr maps. Black, red, blue symbols correspond to $|b_{\rm cut}|> 20^\circ, 30^\circ, 45^\circ$. The filled and open symbols correspond to $30^\circ\leq l_{\rm Gal}\leq 330^\circ$ and $45^\circ\leq l_{\rm Gal}\leq 315^\circ$.
}
\label{fig:fig6_appx}
\end{figure}
Fig. \ref{fig:fig6_appx}, left displays the histograms in the time-differenced maps with the Ecliptic Plane masked out. The source-clipping mask from the full 13-year datasets has been applied and the sky is kept at $30^\circ\leq l_{\rm Gal}\leq 330^\circ$ and $b_{\rm Gal}\geq 20^\circ, 30^\circ,45^\circ$ for the black/blue/red colors. The standard deviations are shown with vertical solid lines corresponding to $\sigma_{F,(A-B)/2}\sim 1.2\times10^{-7}$ GeV/cm$^2$/s/sr.

Fig. \ref{fig:fig6_appx}, right plots the correlation coefficient, $R_{(A-B)\times{\rm 13yr}}$, between the $(A-B)$ and the final 13 yr maps. The figure demonstrates negligible to null correlations implying that the two maps are independent with the dipole power from the $(A-B)/2$ contributing in quadrature to the signal shown in Fig. \ref{fig:fig3}. The precise amplitude, and sign, of the negligible cross-correlation appears sensitive to the (small) addition of $\gamma$-ray photons at higher $E$.

\begin{figure}[h!]
\hspace{0.5cm}\includegraphics[width=4.5in]{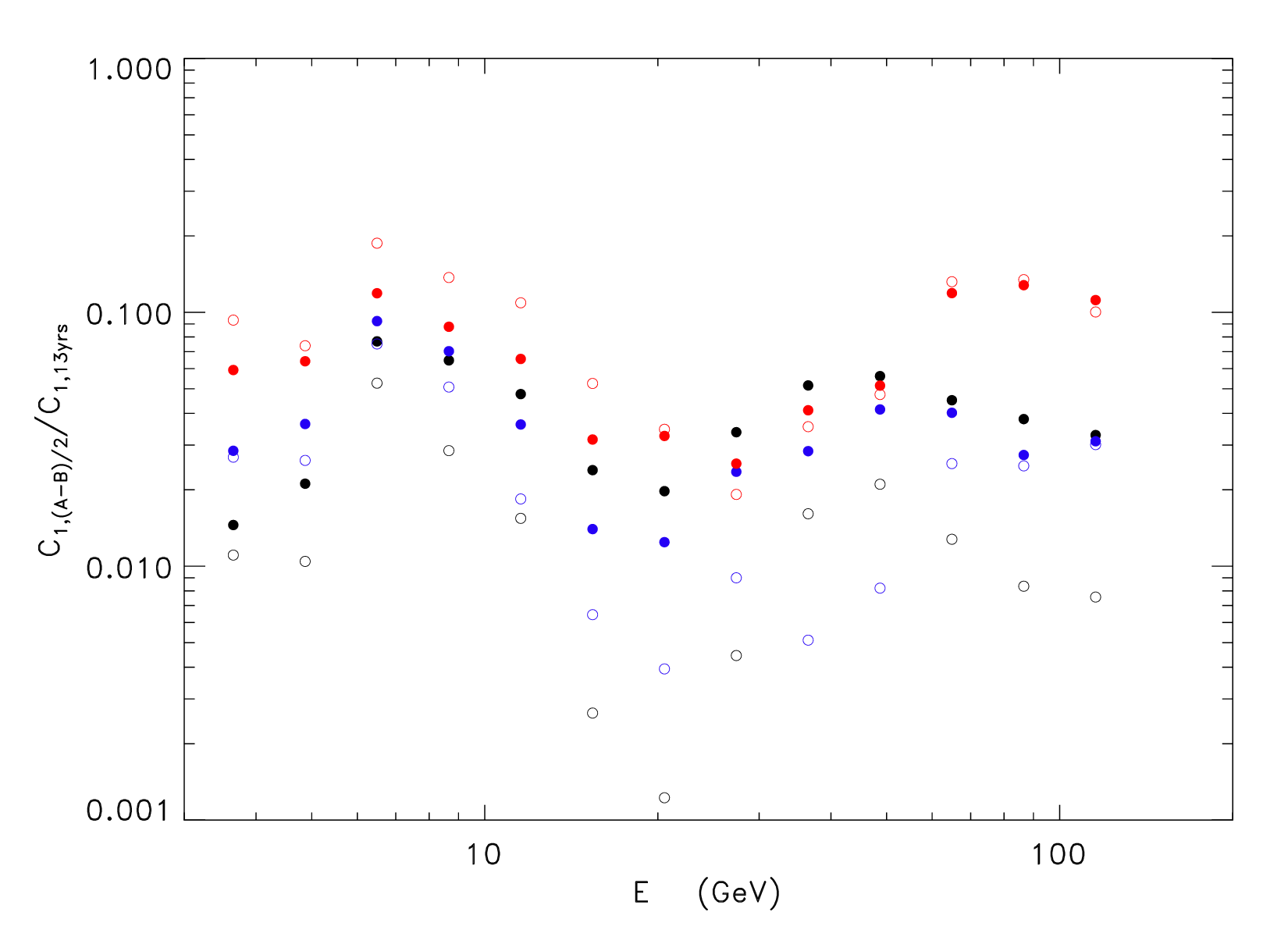}
\caption{\small The ratio of the dipole power, $C_1\equiv \langle d_i^2\rangle$, in the $(A-B)/2$ to 13 yr maps. Open circles correspond to $45^\circ\leq l_{\rm Gal}\leq 315^\circ$  and  filled to  $30^\circ\leq l_{\rm Gal}\leq 330^\circ$. Black, blue, red denote $|b_{\rm Gal}|\geq 20^\circ, 30^\circ,45^\circ$.
}
\label{fig:fig11_appx}
\end{figure}

Fig. \ref{fig:fig11_appx} shows the ratio of the dipole powers, $C_1\equiv \langle d_i^2\rangle$, in the $(A-B)/2$ to that in the 13 yr maps. With the uncorrelated time-varying contribution to the 13 yr dipole added in quadrature 
this implies that the dipole power contributed, $C_1$, from the time-differenced maps is well below ${<\atop\sim} 10\%$. Whereas the 13 yr maps appears to have a stable dipole for the various considered configurations, the $(A-B)/2$ dipole power varies drastically, by more than 2 orders of magnitude, further suggesting the absence of their substantive contributions to the uncovered dipole in Fig. \ref{fig:fig3}. 

Thus the time-variations between the two 6 yr subsets cannot make appreciable contribution to the dipole power found here.

\section*{Acknowledgments}
This work was supported by NASA ROSES 2020 Grant Number 80NSSC20K1597, ``Cosmological Dipole of the Extragalactic Gamma-Ray Background with Multi-Year Fermi LAT Data". 
F.A.B. acknowledges financial support from Grants
PID2021-122938NB-I00 funded by MCIN/AEI/10.13039/501100011033
and by AERDF A way of making Europe" and SA083P17 funded by the Junta de Castilla y Le\'on. We thank Rick Arendt for careful inspection of the final maps.


\end{document}